\documentclass[a4paper,11pt]{article}
\usepackage[T1]{fontenc}
\usepackage[latin1]{inputenc}
\usepackage{float}
\usepackage[dvips]{graphicx}
\usepackage{amssymb}
\usepackage{setspace}

\begin{document}

\title{Dynamic behavior of the interface of strip-like structures in 
driven lattice gases.}

\author{Gustavo P. Saracco and  Ezequiel V. Albano}
\date{}
\maketitle
\emph{Instituto de Investigaciones Fisicoquímicas Teóricas y Aplicadas 
(INIFTA), UNLP, CCT La Plata - CONICET, Casilla de Correo 16 Sucursal 4 
(1900) La Plata, Argentina.}

\begin{abstract}

In this work, the dynamic behavior of the interfaces in both the standard and random driven lattice gas models (DLG and RDLG respectively) is investigated via numerical Monte Carlo simulations in two dimensions. These models consider a lattice gas of density $\rho=1/2$ with nearest-neighbor attractive interactions between particles under the influence of an external driven field applied along one fixed direction in the case of the DLG model, and a randomly varying direction in the case of the RDLG model. The systems are also in contact with a reservoir at temperature $T$. Those systems undergo a second-order nonequilibrium phase transition between an ordered state characterized by high-density strips crossing the sample along the driving field, and a quasi-lattice gas disordered state. For $T\lesssim T_c$, the average interface width of the strips ($W$) was measured as a function of the lattice size and the anisotropic shape factor. It was found that the saturation value $W^{2}_{sat}$ only depends on the lattice size parallel to the external field axis $L_y$ and exhibits two distinct regimes: $W^{2}_{sat}\propto \ln$ $L_y$ for low temperatures, that crosses over to $W^{2}_{sat}\propto L_y^{2\alpha_I}$ near the critical zone, $\alpha_I=1/2$ being the roughness exponent of the interface. By using the relationship $\alpha_I=1/(1+\Delta_I)$, the anisotropic exponent for the interface of the DLG model was estimated, giving $\Delta_I\simeq 1$, in agreement with the computed value for anisotropic bulk exponent $\Delta_B$ with a recently proposed theoretical approach. At the crossover region between both regimes, we observed indications of bulk criticality. The time evolution of $W$ at $T_c$ was also monitored and shows two growing stages: first one observes that $W \propto \ln$ $t$ for several decades, and in the following times one has $W\propto t^{\beta_I}$, where $\beta_{I}$ is the dynamic exponent of the interface width. By using this value we estimated the dynamic critical exponent of the correlation length in the perpendicular direction to the external field, giving $z_{\perp}^I\approx 4$, which is consistent with the dynamic exponent of the bulk critical transition $z_{\perp}^B$ in both theoretical approaches developed for the standard model. A similar scenario was also observed in the RDLG model, suggesting that both models may belong to the same universality class.

\end{abstract}


\section{Introduction}
\label{1}
The study of far-from-equilibrium systems has attracted increasing interest
in the last years \cite{beate} -\cite{marr}, mainly due to the lack of theoretical frameworks, unlike their equilibrium counterparts. However, useful approaches to the understanding of these systems have been developed, like, for example, the analysis of simple models by means of different techniques, such as field-theoretical methods, numerical solving of mean-field equations,
Monte Carlo simulations, dynamic scaling, etc. By using these procedures it is expected that one will establish
and build more general methods to deal with these complex systems. Among the simple
models studied, we shall focus our attention on the driven lattice gas
model (DLG) introduced by Katz, Lebowitz and Spohn \cite{kls}. The study of the DLG model has attracted growing attention due to its interesting far-from-equilibrium 
behavior. This interacting lattice gas, driven into nonequilibrium 
steady states (NESS) by an external field, exhibits remarkable 
properties such as its non-Hamiltonian nature, the violation of the 
fluctuation-dissipation theorem, the occurrence of anisotropic
critical behavior \cite{beate}, the existence of a unique relevant length 
in the anisotropic pattern formation at low temperatures and the consequent
self-similarity in the system at different evolution times \cite{pablo}, etc.\\
The bidimensional DLG model is defined on the square 
lattice assuming a rectangular geometry ($L_{x} \times L_{y}$) and
using periodic boundary conditions. Lattice configurations are defined 
by means of occupation numbers $n_{i,j}=1$ or $0$ for each site of coordinates
$(i,j)$ depending on whether the site is occupied or empty, respectively. By assuming nearest-neighbor (NN) attractive interactions between particles of the gas ($J > 0$) and in absence of any driven field,
the Hamiltonian (H) corresponds to the Ising lattice gas 

\begin{equation}
H =  -4J \sum_{<ij,i'j'>} n_{i,j} n_{i',j'}    ,
\label{Hamil}
\end{equation}

\noindent where the summation runs over NN sites only. By considering a
driven field $E$, the coupling to a
thermal bath at temperature $T$ is accounted by a modified Metropolis rate, given by

\begin{equation}
min [1, exp (\Delta H - \eta E/ k_{b}T) ]      ,
\label{Metro}
\end{equation}

\noindent where $k_{b}$ is the Boltzmann constant, $\Delta H$ is the
change in $H$ after the attempted particle movement, and 
$\eta = [-1, 0, 1]$ for a particle hopping 
(against, orthogonally, along) the field. Also, the temperature $T$ of the reservoir in contact with the sample 
is reported in units of $J/k_{b}T$.

It is well known that for high enough temperatures the DLG model
exhibits a lattice-gas-like disordered state. However, at low
temperatures anisotropic NESS emerge. This ordered phase is
characterized by the presence of strips of high particle density
crossing the lattice in the direction parallel to the external field.
So, for half-density of particles and at a well-defined critical bulk
temperature $T_{c}$, the DLG model undergoes a second-order phase 
transition \cite{beate}. Extensive Monte Carlo simulations have shown that 
$T_{c} \simeq 1.41 T_{c}^{O}$, where $T_{c}^{O} = 2.269$ $J/k_{b}$ 
is the Onsager critical temperature of the Ising model \cite{beate}.

The issue of the universality class of the DLG model has become the
subject of a long-standing debate. On the one hand, the first nonlinear Langevin equation for this model was
developed by Janssen \textsl{et al.}, where it is argued that the main nonlinearity is the current term 
generated by the external field \cite{JS}. By working out this equation a set of critical exponents was calculated 
and a new universality class was found, which is different from the universality class of the Ising model. 
The values of the critical exponents were early confirmed by several numerical simulations \cite{ktl}. 
Later on, another Langevin equation was developed for a variant of the DLG model, called random DLG model (RDLG). 
In this model, the driving field is still applied along one axis as in the DLG model, but its direction 
changes randomly causing the net particle current, which characterizes the DLG model, to vanish. By analyzing this system it was found that its critical behavior falls outside of both the Ising and the DLG universality classes \cite{rkls}.
On the other hand, Garrido \textsl{et al.} \cite{gallegos1} have revisited the subject by using a different approach 
and found a new Langevin equation for the DLG model, in which the main nonequilibrium effect is due to the underlying 
anisotropy. This equation was already known because, in the limit of infinite driving fields 
(the case that commonly is studied), it describes the behavior of the RDLG model. 
According to their results, Garrido \textsl{et al.} have concluded that both DLG and RDLG 
models belong to the same universality class in the limit of an infinite driving field. 
These results have stimulated interesting theoretical discussions \cite{tanos, reply}, 
and recent numerical results  \cite{gallegos2, alsa} agree with the universality class of the 
Langevin equation derived from this last approach. However, subsequent numerical simulations and 
theoretical work may suggest that the DLG model belongs to the class of the Langevin equation 
early developed by Janssen \textsl{et al.} \cite{gamba} and that the RDLG model has a universality 
class of its own \cite{gamba2}.\\

Besides the bulk phase transition, another interesting feature of the DLG model is the interfacial behavior 
and the possible existence of a roughening transition. This basically consists in the interface change from flat 
(or smooth) to rough when the temperature increases \cite{bara}. Originally studied in crystal growth models, it 
has been theoretically characterized by  a Hamiltonian proportional 
to the interface surface (the proportionality constant being the \textsl{stiffness} or \textsl{surface tension}), 
plus a periodic potential that represents the discrete nature of the crystal surface height. 
It has been found that the transition can be mapped to the Kosterlitz and Thouless 
transition \cite{lapu} at a nontrivial roughening critical temperature, $T_R$, where at low 
temperatures ($T<T_R$), the saturated statistical width $W^2_{sat}$ (see Section \ref{2}) is 
independent of the linear system size ($L$) parallel to it. 
Furthermore, for  $T\geq T_R$ one has  $W^2_{sat} \propto \ln$ $L$. 
The existence of the roughening transition has been confirmed in several simulations of solid-on-solid (SOS) 
models, by measuring observables such as the height-height correlation function, the difference of concentration
 in the interface layers, the fluctuations in the number of particles belonging to the interface \cite{weeks},
 etc., and in experimental work concerning different metallic interfaces \cite{lapu}. \\

In the equilibrium Ising model, the existence of a roughening transition at a nontrivial $T_R$ depends on the 
system dimension. So, for $d=2$ one has $T_R=0$ and for $d=3$ one has $T_R\simeq 0.55$ $T_c$  ($T_c$ is the 
critical temperature of the bulk phase transition) \cite{bruk}. Furthermore, in the rough phase ($T>T_R$), the 
structure factor $G(\overrightarrow{q})$ (i.e., the Fourier transform of the height-height correlation 
function $\langle h(\overrightarrow{x})h(0) \rangle-\langle h \rangle^{2}$, with $h(\overrightarrow{x})$ being 
the interface height given at the position $\overrightarrow{x}$), behaves as 
$G(\overrightarrow{q})\propto1/\overrightarrow{q}^{2}$  
in the long-wavelength limit, $\overrightarrow{q}$ being  the $d-1$ dimensional wave 
vector of the capillary waves \cite{vanB}. This leads to a divergence in $W^{2}_{sat}$ for the spatial dimension $d\leq3$. In particular, for the bidimensional Ising model it has been found that $W^{2}_{sat}\sim$ $L^{p}$, with $p=1$ \cite{bab}.\\ 

For the case of the DLG model in $d=2$, the investigations have been directed to study the effects of 
the external drive on this kind of transition. 
Regrettably, neither the theoretical approaches \cite{leung, mon} nor numerical simulations arrived 
at definitive results. The simulations performed by Leung \textsl{et al.} on the DLG model 
at low temperatures and several values of the external drive $E$ have shown that the value of 
the exponent $p$ is 
consistent with $p\rightarrow0$ when $E$ increases, indicating the suppression of roughness, in contrast to the equilibrium Ising model \cite{mon}. Measurements of the structure factor of the height-height correlations are consistent with this picture, revealing that the singular behavior of $G(\overrightarrow{q})$ is less severe than 
in the Ising model, although these data are affected by finite-size effects and it is no longer possible 
to give a more quantitative conclusion \cite{mon}. Subsequently, new measurements of $G(\overrightarrow{q})$ 
performed by Leung \textsl{et al.} using larger square lattices (the lattice size was up to 600 lattice units) 
showed that $G(\overrightarrow{q})\propto1/\overrightarrow{q}^{2-\eta}$, 
$\eta=1.33$, for small $\overrightarrow{q}$ and both $E=2$ and $E=50$ \cite{leung2} at $T\backsimeq 2.05 J/k_b$ 
$\ll T_c$. If this behavior persists in the $\overrightarrow{q}\rightarrow0$ limit, 
it would be consistent with a severe roughness supression, due to an enhacement of the surface 
tension on large length scales \cite{leung2}. In the same work, Leung \textsl{et al} 
also analyzed the interface behavior of the RDLG model in lattices of sizes up to 800 
lattice units and at the same temperature. They found a less clear behavior than in the DLG model. 
In spite of the large size used and the noisy data near $\overrightarrow{q}=0$, Leung \textsl{et al.} 
have concluded that $G(\overrightarrow{q})$ crosses over from $G(\overrightarrow{q})\propto 1/q$ to 
$G(\overrightarrow{q})\propto q^{-x}$ with $x<1$ in the limit $\overrightarrow{q}\rightarrow0$, suggesting 
again the suppression of roughness \cite{leung2}. From the theoretical point of view, some progress has been made only in the RDLG model, since this model has no current term in the equation of motion. 
It has been found that $W^2_{sat}$ $\propto$ $\ln$ $L$, also consistent with $p=0$ \cite{zialog}. However, 
this dependence is not consistent with the simulation results for the short-wavelength limit, 
because a logarithmic increase of the width would correspond to a behavior of the 
type $G(\overrightarrow{q})\propto 1/q$ in the $q\rightarrow0$ limit.

In view of the outlined controversies and the lack of some conclusive 
answers to existing open questions, the aim of this paper is to provide 
an alternative evaluation of the dynamic critical exponents that describe 
the second-order phase transition of the DLG model, complementary to the one 
performed in ref. \cite{alsa}, and to investigate the possible existence of a roughening transition. 
The attention will be focused on the properties of the interface between the high particle density strips 
and the gas phase, monitoring the behavior of observables such as the interface width, and the bulk and interface 
energies together with their fluctuations. Our study is based on extensive numerical Monte Carlo simulations 
and the interpretation of the obtained data is achieved by using dynamic scaling theories. We also performed 
simulations of the RDLG model in order to compare its behavior with that of the DLG model. \\
This paper is organized as follows: in Section \ref{2} the theoretical background is summarized, in Section 
\ref{3} we give a brief description of the model; in Section \ref{4} we present and discuss our results, and finally in Section \ref{5} our conclusions are stated.

\section{Dynamic Scaling Approach for the Interface Roughness of the 
Strips}
\label{2}
\begin{figure}[H]
\centering
\includegraphics[height=8cm,width=11cm,clip=,angle=0]{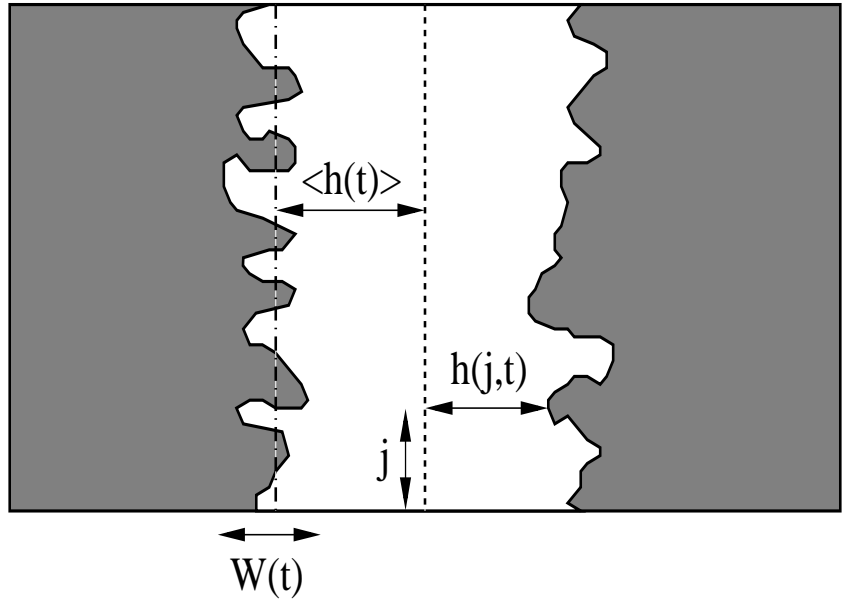}
\caption{Sketch of a strip of the DLG model (white region) showing the  
relevant quantities used for the calculations: the distance from the 
center 
of the strip (indicated by the dashed line) up to the interface at the position of the $j^{th}$ 
column at time $t$ given by $h(j,t)$; the average position of the 
interface
at time $t$, $\langle h(t) \rangle$ (indicated by the dot-dashed line) (eq. (\ref{hmediodef})); and the average width of the interface at time 
$t$,
$W(t)$, given by eq. (\ref{anchodef}). The gray regions represent the gas phase, and the external field is applied along the vertical axis.}
\label{Picture2}
\end{figure}

In order to characterize the dynamic evolution of the interfaces in the DLG model, we 
first measure the \textit{average interface position} (see figure \ref{Picture2}),  given by

\begin{equation}
\langle h(t) \rangle = \frac{1}{N_I} \sum_{j=1}^{N_I} h(j,t),
\label{hmediodef}
\end{equation}

\noindent where $h(j,t)$ is the distance measured from the center 
of the strip up to
the interface at the position of the $j^{th}$ column at time $t$, and 
$N_I$ 
is the total number of particles at the interface. Subsequently, the \textit{interface width} 
or \textit{roughness} is defined as the rms of the mean  
interface position 

\begin{equation}
W(L_y,t) = \sqrt{\langle h(t)^2 \rangle -\langle h(t) \rangle^2}.
\label{anchodef}
\end{equation}

\noindent Notice that the summation in both quantities runs over all the particles of the interface because in general
we have $N_I\geq L_y$ (i.e., the number of columns given by the lattice size)  and in consequence $h(j,t)$ is not a single-valued function of the j-coordinate due to the presence of overhangs and/or  bubbles. In the simulations we will compute $W$ as a function of time, at different temperatures and using several lattice sizes, in order to study the 
change of roughness due to the evaporation/condensation of particles at the 
strip surfaces.  Due to the selected initial condition, a single strip at $T=0$, the 
interface has zero width at $t = 0$.

\begin{figure}[H]
\centering
\includegraphics[height=14cm,width=9cm,clip=,angle=-90]{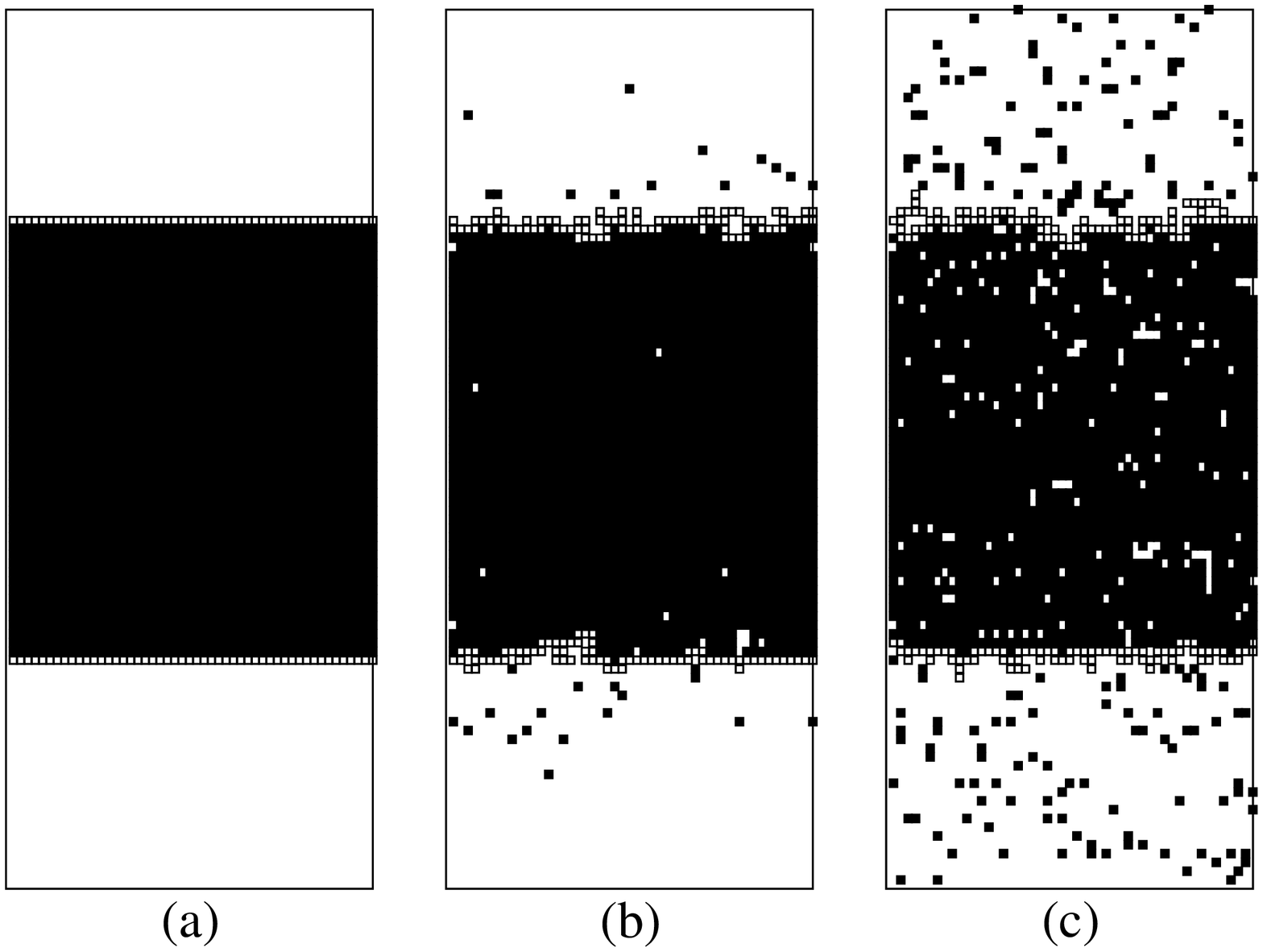}
\includegraphics[height=14cm,width=9cm,clip=,angle=-90]{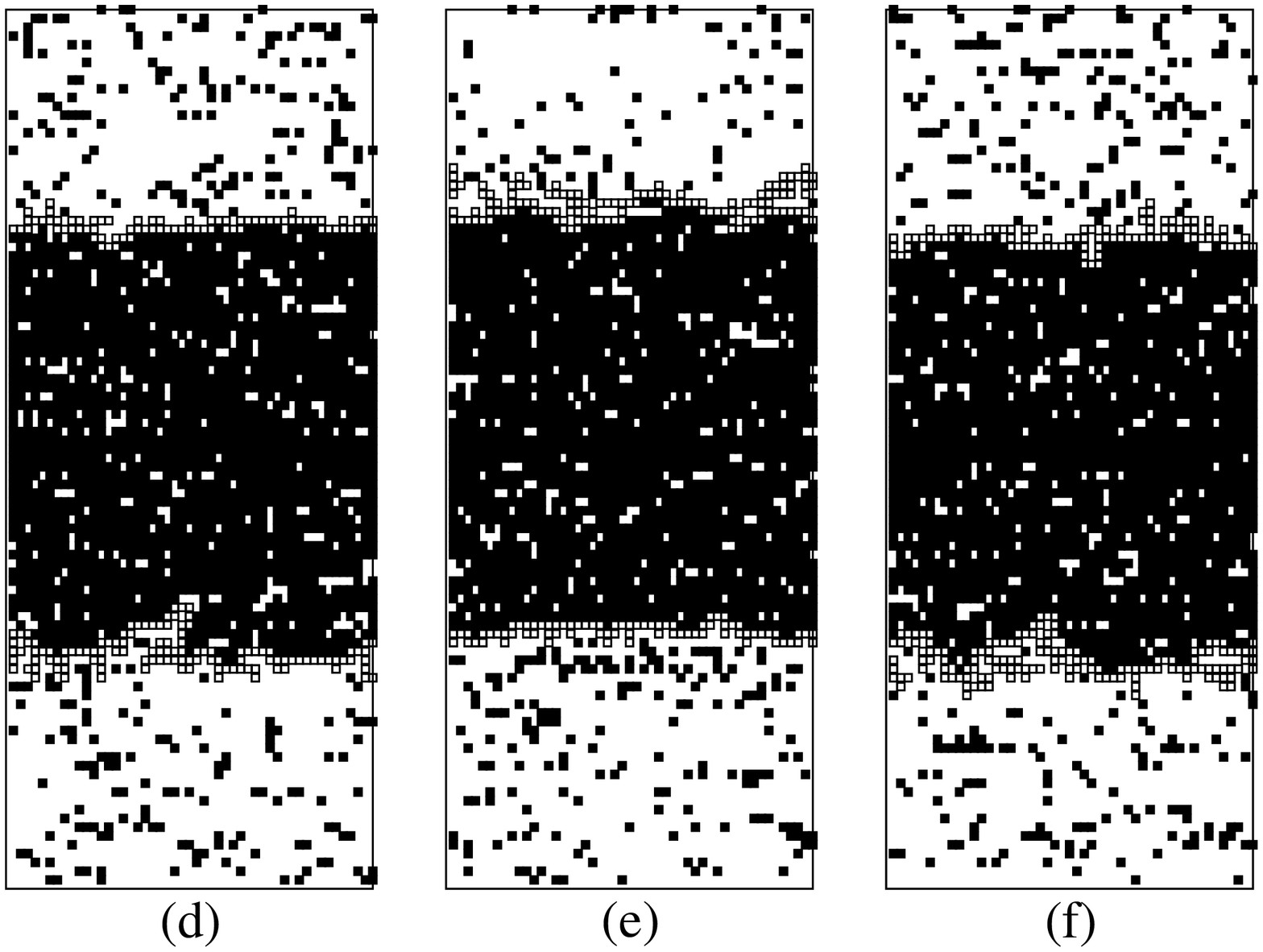}
\caption{Snapshot configurations corresponding to the IDLG model.
Figure (a) shows a single strip containing all particles, corresponding to the 
ground
state configuration ($T = 0)$ that is used as initial condition in all  
simulations reported in this work i.e. at t=0. Figures (b)-(f) show the subsequent time 
evolution of 
the strip when the system is at $T = 2.6 < T_{c}$. 
The lattice size is  $L_x = 100$, $L_y = 50$ (the field is applied along the horizontal axis). Particles are shown in black and empty sites are left in white. Empty squares are used to show the 
interfaces 
of the strip. From left to right, the corresponding evolution times at which the snapshots were taken are the following: (a) $t=0$, (b) $t=100$, (c) $t=1024$, (d) $t=10000$, (e) $t=100489$, and (f) $t=10^6$, measured in Monte Carlo time steps (MCS). }
\label{Picture1}
\end{figure}

In the DLG model, the properties and even the existence of interfaces 
depend on the temperature $T$. In fact, for $T \geq T_c$, 
the system is in the disordered phase
and there are no interfaces at all.  However, for $T < T_c$, interfaces 
are present due to the strip-like patterns characteristic of the ordered 
phase (see figure \ref{Picture1}).  If the initial configuration of the system consists of a single strip placed at the center of the lattice with no holes and two flat interfaces (see Section \ref{3}) and it is taken to evolve  at a temperature  $T > 0$, the particles at the interfaces may leave the strip diffusing into the gas-like  phase. The inverse process also takes place, and some particles of the gas-like phase may stick again at the interfaces.  Moreover, there are two diffusion processes at the interface: one is driven by the
external field, which sets a current along the interface, while
the other corresponds to the diffusion of particles in the direction
perpendicular to the driving field toward the bulk of the strip. These processes  
changes the interface width $W$ (eq. (\ref{anchodef})). These changes in $W$ come from both the intrinsic width of the interface, $W_{in}$ which is of the order of the correlation width of the bulk correlation length, and the thermal fluctuations of the local mean position of the interface, i.e. of the capillary waves \cite{mon}. If the system is in the rough phase and far from the critical point,  $T\ll T_c$, the intrinsic width is negligible and the interface evolution is governed by the fluctuations of the capillary waves. On the other hand, if $T\lesssim T_c$, the bulk correlation length becomes relevant, the intrinsic width is no longer negligible, and $W\propto W_{in}$. \\

In order to describe the dynamic scaling properties of the interfaces 
of the strips of the DLG model, we used the scaling Ansatz early 
proposed by Family and Vicsek \cite{bara,famvis}. So, it is assumed 
that the time evolution of roughness obeys the following scaling law

\begin{equation}
W(L_y,t) \sim L_y^{\alpha_I} f\Bigg{(}\frac{t}{L_y^{z_{\parallel}^I}} \Bigg{)},
\label{fvis2}
\end{equation}

\noindent where $f(u)\propto u^{\beta_{I}}$ for $u \ll 1$ and
$f(u)\rightarrow {\it constant}$ for $u \gg 1$, with $z_{\parallel}^I=\alpha_I
/\beta_{I}$. The limit $u \rightarrow 1$ sets the crossover time
$t_{x}\sim L_y^{z_{\parallel}^I}$ between both regimes, $u \ll 1$  and $u\gg1$,
respectively. Furthermore, $\alpha_I$, $\beta_{I}$, and $z_{\parallel}^I$ are defined as 
the roughness, growth, and dynamic exponents, respectively. We used the subscript or 
superscript ``I'' in order to distinguish these exponents from those 
critical exponents of the second-order phase transition of the DLG model.
The dynamic exponent $z_{\parallel}^I$ describes the time evolution of
the interface correlation length along the direction parallel to the interface according to 
$\xi_{\parallel}^I \propto t^{1/z_{\parallel}^I}$ \cite{bara}. 
Thus, for a finite sample of size $L_y$ and $t \rightarrow \infty $ the 
correlation length remains finite ($\xi_{\parallel}^{I}\simeq L_y$) and the
interface width reaches a saturation value that depends on the system
size as $W_{sat}(L_y)\sim L_y^{\alpha_I}$. Since we want to relate the interface behavior with the critical behavior of the system, we expect that equation (\ref{fvis2}) will hold for temperatures near the critical point ($T\lesssim T_c$).   
\newpage
\newpage

In order to adapt the dynamic scaling of Family and Vicsek (eq. (\ref{fvis2})) to the DLG model, 
we have to recall that the system is intrinsically anisotropic and, consequently, the exponents entering  
into equation (\ref{fvis2}) must be consistent with that anisotropy. 
As we will see below, the width of the interface saturates 
for $T \lesssim T_{c}$ due to finite-size effects. So, in the limit $u\gg 1$, equation (\ref{fvis2}) becomes

\begin{equation}
\label{wsatkls} 
W_{sat}(T)\sim L_{y}^{\alpha_{I}} \qquad T\lesssim T_c,
\end{equation}

\noindent where $\alpha_{I}(T)$ is the roughness exponent, and $L_{y}$ is the lattice size along the direction of the external field.

\noindent For $T = T_{c}$ and in the $L_y\rightarrow \infty$ limit, the interface correlation length in the perpendicular direction of the field axis $\xi_{\perp}^I$ always diverges with time  as a power law with dynamic exponent $z_{\perp}^I$ \cite{bara}. As we will see later, the dynamic evolution of $W$ at $T=T_c$ in large systems (see next Section) shows two different growing regimes (see figure \ref{regikls}). In the first regime $W$ diverges logarithmically, i.e., $W \propto \ln t$, while at later times $W$ diverges as a power law, $W \propto t^{\beta_{I}}$. So, according to the standard dynamic scaling theory \cite{hav}, we have 

\begin{equation}
\label{wcreckls} W\sim \xi_{\perp}^I\propto t^{1/z_{\perp}^I} \qquad
T\simeq T_c.
\label{diverperp}
\end{equation}

\noindent Since we are working with $T\lesssim T_c$, the major contribution to $W$ is due to the intrinsic width, so $W_{in}\sim \xi_{\perp}^B \sim \xi_{\perp}^I$, where  $\xi_{\perp}^B$ is the bulk correlation length in the perpendicular direction  \cite{mon}. Comparing equations (\ref{fvis2}) and (\ref{diverperp}) for the limit 
$u\ll 1$, it follows  that $\beta_{I}=1/z_{\perp}^I$. 

\noindent In the range $0<T<T_c$, the occurrence of a crossover time $t_x$ (see eq. (\ref{fvis2})) is justified by the fact that the correlation length in the parallel direction to the applied field ($\xi_{\parallel}^I$), calculated from the height-height correlation function $\langle h(\overrightarrow{x})h(0) \rangle-\langle h \rangle^{2}$,  grows, and since the system size is finite in the $L_y$-direction, the saturation effects are observed  when $\xi_{\parallel}^I \sim L_{y}$ for $t\sim t_x$. 
Recalling that $\xi_{\parallel}^I\propto t^{1/z_{\parallel}^I}$ we have
that

\begin{equation}\label{tcrosskls}
\xi_{\parallel}^I(t_x)\sim L_{y}\propto
t_{x}^{1/z_{\parallel}^I}.
\end{equation}

\noindent Consequently, 
from the relation $z_{\parallel}^I=\alpha_I/\beta_{I}$ that holds for the Family-Viscek
dynamic Ansatz \cite{famvis}, we have

\begin{equation}
\label{zbdkls}
z_{\parallel}^I=\frac{\alpha_I}{\beta_{I}}=\alpha_I\, z_{\perp}^I.
\end{equation}

\noindent From the scaling theory developed for the DLG model \cite{beate}, 
it is well known that the dynamic exponents governing the bulk critical behavior, $z_{\parallel}^B$
and $z_{\perp}^B$, are related through the anisotropic exponent
$\Delta_B$ as follows \cite{beate}

\begin{equation}
\label{zbeate} 
z_{\parallel}^B=z_{\perp}^B/(1+\Delta_B).
\end{equation}

\noindent We will relate these last two equations by assuming that the interface exhibits the same anisotropic behavior. So, the anisotropic exponent for the interface can be defined in the following way:
\begin{equation}
\label{alfadelta} 
\alpha_I=\frac{1}{1+\Delta_I}.
\end{equation}

The above assumptions were based on previous studies of equilibrium models where a conservation law is present (model B), and also in the DLG model. In fact, the bulk modes are slow and couple to the interface modes affecting its behavior \cite{ turski,kawa}. In the DLG model, it has been found that the bulk modes are also slow and decay slower than the interface modes, suggesting that they take part in the interfacial behavior \cite{leung,yeung}. By virtue of these findings, is is expectable that the anisotropy effects induced by the driving field will be also present on the interface, so that $\Delta_B=\Delta_I$. In this way, by measuring both $z_{\perp}^I$ and $\alpha_I$ it would be possible to
determine the critical dynamic exponent $z_{\parallel}^I$ by using 
equation (\ref{zbdkls}). Furthermore, we can also compute the anisotropic exponent $\Delta_I$ 
by using the determined value of $\alpha_I$. It is worthwhile remarking
that $\Delta_B \simeq 1$  ($\Delta_B =  2$) has been predicted theoretically by 
the alternative coarse-grained description of the model developed by
Garrido \textsl{et al} \cite{gallegos1}, (by the coarse-grained field theoretical equation developed by Janssen \textsl{et al}
\cite{JS}). Consequently, the measurement of the interface properties of the DLG model close to criticality is expected to be
helpful in clarifying the issue of the universality class of the 
DLG model.

\section{The Simulation Method}
\label{3}

The DLG and  RDLG models are simulated in square lattices of size
$L_{x}\times L_{y}$, where $L_{x(y)}$ 
is the lattice length in the perpendicular (parallel) direction 
to the applied drive. In order to investigate the behavior of the interface at low temperatures, simulations were performed by using lattices of sizes within the ranges $62\le L_x \le 284$ and $20\le L_{y}\le 360$
lattice units (LU), keeping the shape factor 
$S=L_y^{\nu_{\perp}/\nu_{\parallel}}/L_x=0.0786$ fixed. For setting this value we employed the critical exponents determined by Garrido \textsl{et al.} \cite{gallegos1}, i.e., $\nu_{\perp}\simeq0.63$ and $\nu_{\parallel}\simeq1.22$. Furthermore, we also performed a second set of simulations keeping $L_y=20$ LU fixed but changing $L_x$ in the range $40\le L_x \le 4000$, in order to study the dependence of the interface behavior on $L_x$. Notice that $S$ is not fixed at all for these lattices.\\   
In the case of simulations at the critical
temperature, we employed a larger lattice with $L_x = 346$ and 
$L_y = 524$ LU in order to avoid the early occurrence of finite-size effects. 
In all cases the particle density is fixed at $\rho_{0} = 1/2$, 
so that the DLG model exhibits a second-order phase transition \cite{beate}. 
The magnitude of the driving field is 
kept constant at $E=50$ $J$ for all simulations. This value implies, for practical purposes, that we are considering the
$E \rightarrow \infty$ limit (see equation (\ref{Metro})). So, the studied models
become the infinite field DLG model (IDLG) and the infinite random field (IRDLG) model. The Monte Carlo time step (MCS) is defined such as all sites ($L_x \times L_y$) of the lattice are selected once (on the average) in order to attempt the displacement of a particle to an empty neighbor site.\\
The temperature is measured in units of $J/k_{b}$, $k_{b}$ being the Boltzmann
constant. Under these conditions, the bulk critical temperatures of both
models have been estimated in previous work \cite{beate, gallegos2,alsa}, namely,
$T_{c}(IDLG) \cong 3.20$ and $T_{c}(IRDLG)\cong 3.16$. We monitored the time evolution of $W(t)$ (eq. (\ref{anchodef})) over many realizations of the experiment, typically 100-3000 realizations according to the temperature and lattice sizes.\\
The initial condition used in order to start the simulations was the ground state configuration (GSC), which is the configuration that the system has at $T=0$. It consists of a single strip, free of defects and placed at the
center of the lattice (see figure \ref{Picture1}(a)). Subsequently, the 
time evolution of the interface is followed, measuring its position $h$ with respect to the center of the lattice (equation (\ref{hmediodef})) and its width $W$ (equation (\ref{anchodef})) by identifying the particles belonging to the interface (this process will be described below). From the time evolution of $W$ we measure the saturation value $W_{sat}$ by averaging over a time interval long after the crossover time given by equation (\ref{tcrosskls}).\\
 
The identification of the interface is performed by employing the algorithm developed in ref. \cite{vero} for diffusion fronts. Let us now briefly explain how the algorithm works. By taking as an example the snapshot of the system exhibited in figure \ref{Picture1}, we divide the lattice in (horizontal) rows, and suppose that we want to measure the position of the interface that is between the ($L_x/2+1$)-th and ($L_x$)-th rows (recall that in figure \ref{Picture1} $L_x$ runs along the vertical axis). Then, the interface identification is performed through a process of three stages. In the first stage, we impose that all sites in the ($L_x/2+1$)-th row are occupied with particles \footnote{If neccessary, all empty sites on this row will become occupied}, and are labelled with an integer number. Then, the lattice is swept sequentially from the ($L_x/2+1$)-th row to the ($L_x$)-th row, and by applying the standard Hoshen-Kopelman algorithm (HK) \cite{hav} to nearest neighbors occupied sites the largest cluster, which corresponds to the strip running along the $L_y$ direction can be identified. In the second stage the HK algorithm is repeated but \textit{in the opposite way}, from the ($L_x$)-th to ($L_x/2+1$)-th rows. In this case, the largest cluster of empty sites, linked by both nearest- and next-nearest- neighboring sites, is determined.\\
At this point, the portion of the lattice we have considered, of size $L_x/2$ $\times$ $L_y$ has two clusters, and the interface can easily be defined as those sites of the largest cluster of occupied sites having at least a neighbor belonging to the largest cluster of empty sites. This is done in the third stage. Once the overall identification process is finished, the interface height, measured from $L_x/2$,  and its width are calculated. One important feature of this method is that takes into account the bubbles and overhangs (they can be observed in the snapshots of figure \ref{Picture1}) responsible for the intrinsic width of the interface, while previous studies neglect them by applying a strong coarsening of the interface, smoothing it by spreading the local density on neigboring sites \cite{mon,leung2}. \\

\section{Results and Discussion}
\label{4}

\begin{figure}[H]
\centering
\includegraphics[height=15cm,width=11cm,clip=,angle=270]{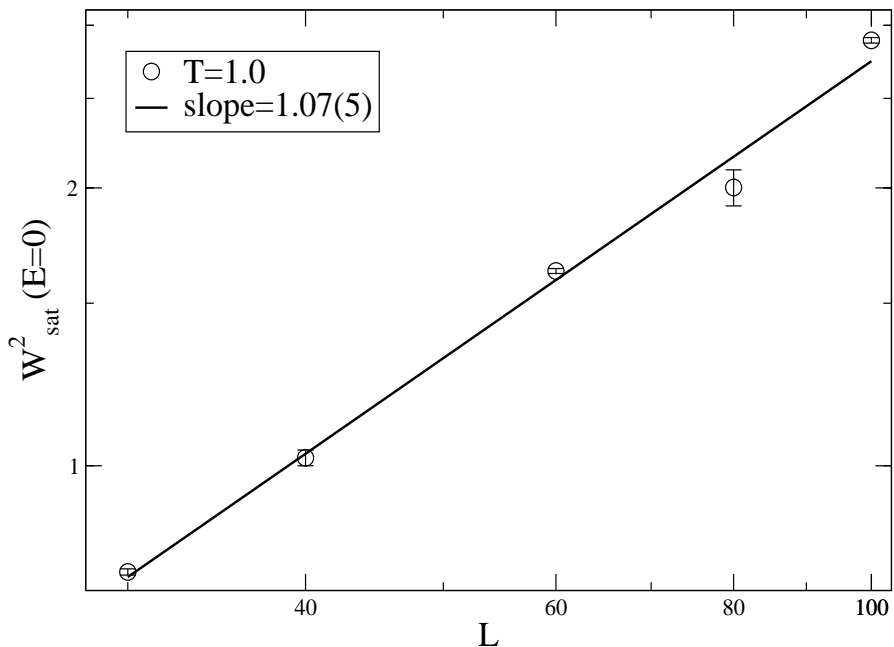}
\caption{Log-log plot of the saturated width $W^{2}_{sat}$ versus $L$ for the Ising model with conserved dynamics at $T=1$ $J/k_b$.}
\label{e0}
\end{figure}

The protocol used to register the time evolution of the interface width(s) is described first. All simulations were performed in the following way: we started the simulations from the GSC configuration and then we let the system to evolve at  a temperature $T\lesssim T_c$ for different lattice sizes $L_x$, $L_y$ that are specified to each case. Then we monitored the time evolution of the roughness $W$ of each interface, stopping the simulations when either the strip no longer percolated along the direction parallel to the driving field, and/or more than one strip were detected. Then a new simulation was performed.\\
In order to test the algorithm used to determine the interface position, we simulated the model at $E=0$, i.e., the Ising model with conserved (i.e., Kawasaki) dynamics. We fixed $T=1$ and used square lattices in the range $20 \leq L_x=L_y=L\leq 100$. As it was mentioned in the Introduction, theoretical arguments performed by Ben Abraham \textsl{et al} \cite{bab} and numerically confirmed by Leung \cite{leung} state that the saturation width behaves as $W_{sat}^{2}\sim L^{p}$, with $p=1$. Figure \ref{e0} shows a plot of the saturation width $W_{sat}^2$ versus $L$. Our best least squares fit of the obtained data gives $p=1.07(5)$, which is in good agreement with the already mentioned expectation. This result allows us to show that the algorithm is reliable for our purposes, since the interface can be well characterized.

\begin{figure}[H]
\centering
\includegraphics[height=15cm,width=11cm,clip=,angle=270]{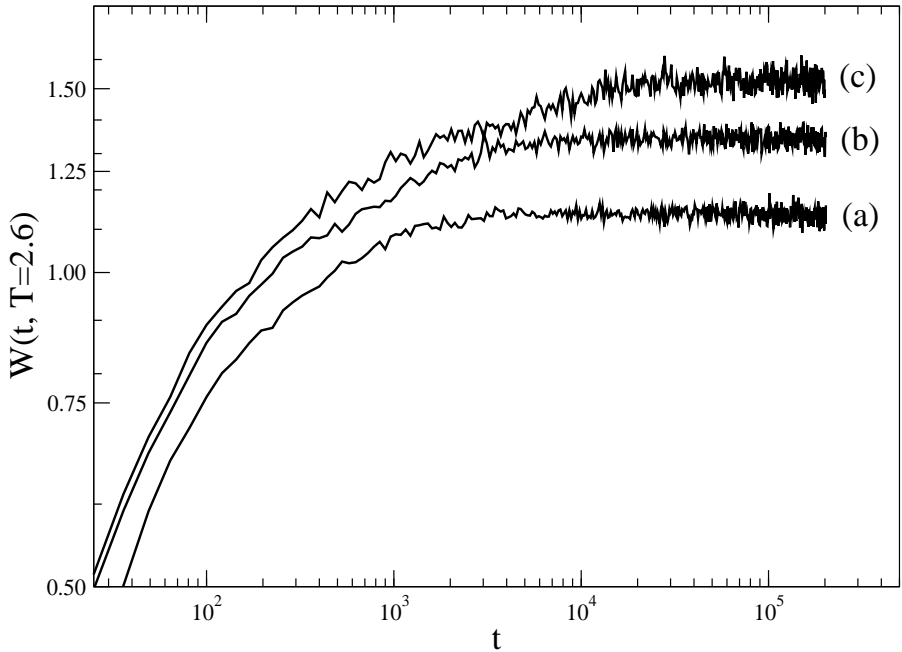}
\caption{Log-log plots of the interface width $W$ versus $t$ measured
at $T = 2.60 < T_{c}$ for the IDLG model. The lattice sizes employed were (a)
$L_x=62$, $L_y=20$, (b) $L_x=100$, $L_y=50$, and (c) $L_x=208$,
$L_y=200$. The curves were obtained by averaging over
2000, 300, and 60 different  realizations in (a), (b), and (c),
respectively.}
\label{wtf}
\end{figure}

Now we focus our attention to the DLG model. In figures \ref{Picture1}(a-f) a typical time evolution of the system from the GSC initial condition is shown. As we can observe in the figures, the strip does not move appreciably with respect to the center of the lattice, so cluster fluctuations due to the transversal periodic boundary conditions are irrelevant. This prevents miscalculations of the interface position. Consequently, the interface width undergoes an initial increase and eventually reaches a saturation value at long enough times. As expected, the saturation value depends on the lattice size, as shown in figure \ref{wtf}. 

\begin{figure}[H]
\centering
\includegraphics[height=12cm,width=9cm,clip=,angle=270]{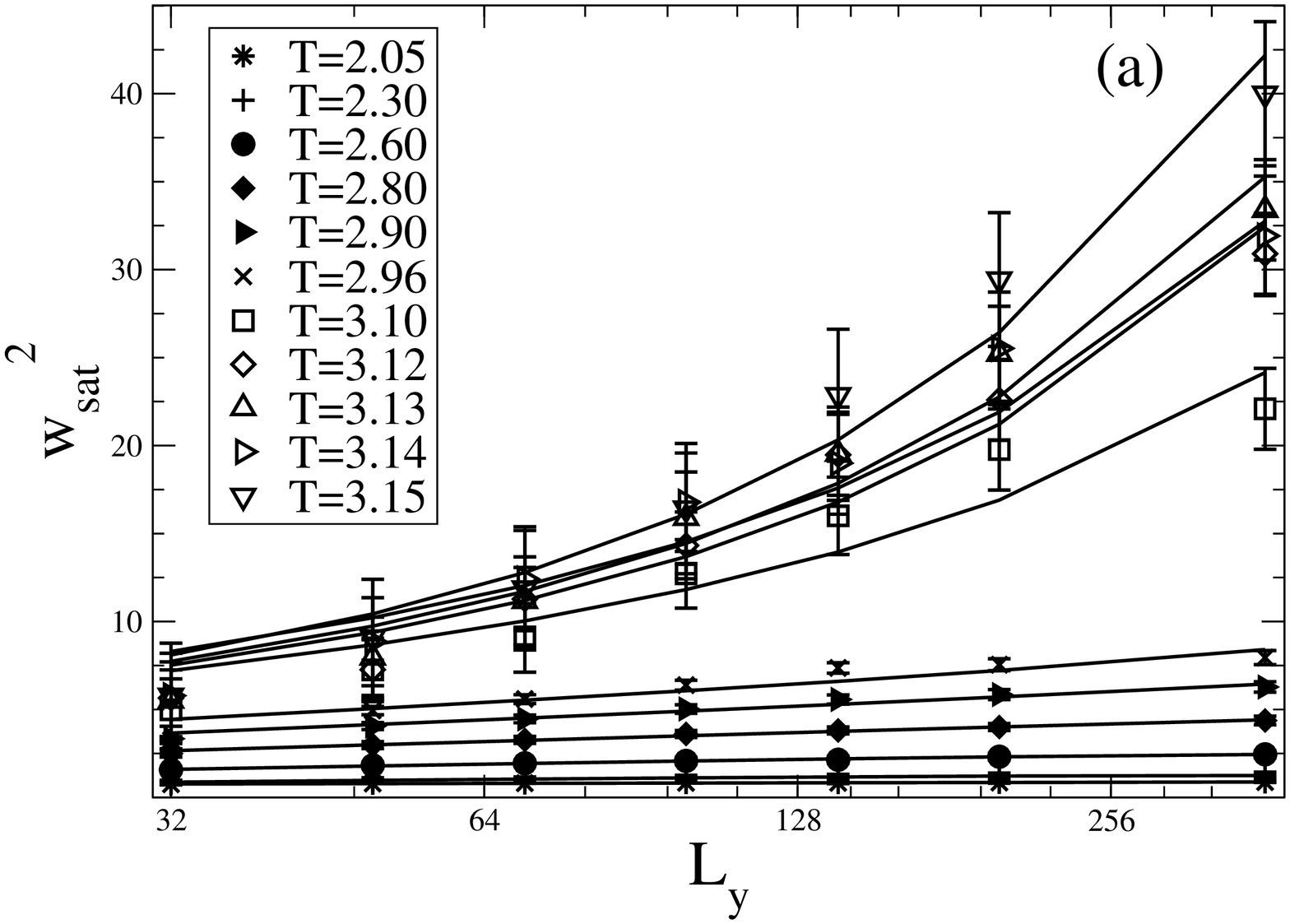}
\includegraphics[height=12cm,width=9cm,clip=,angle=270]{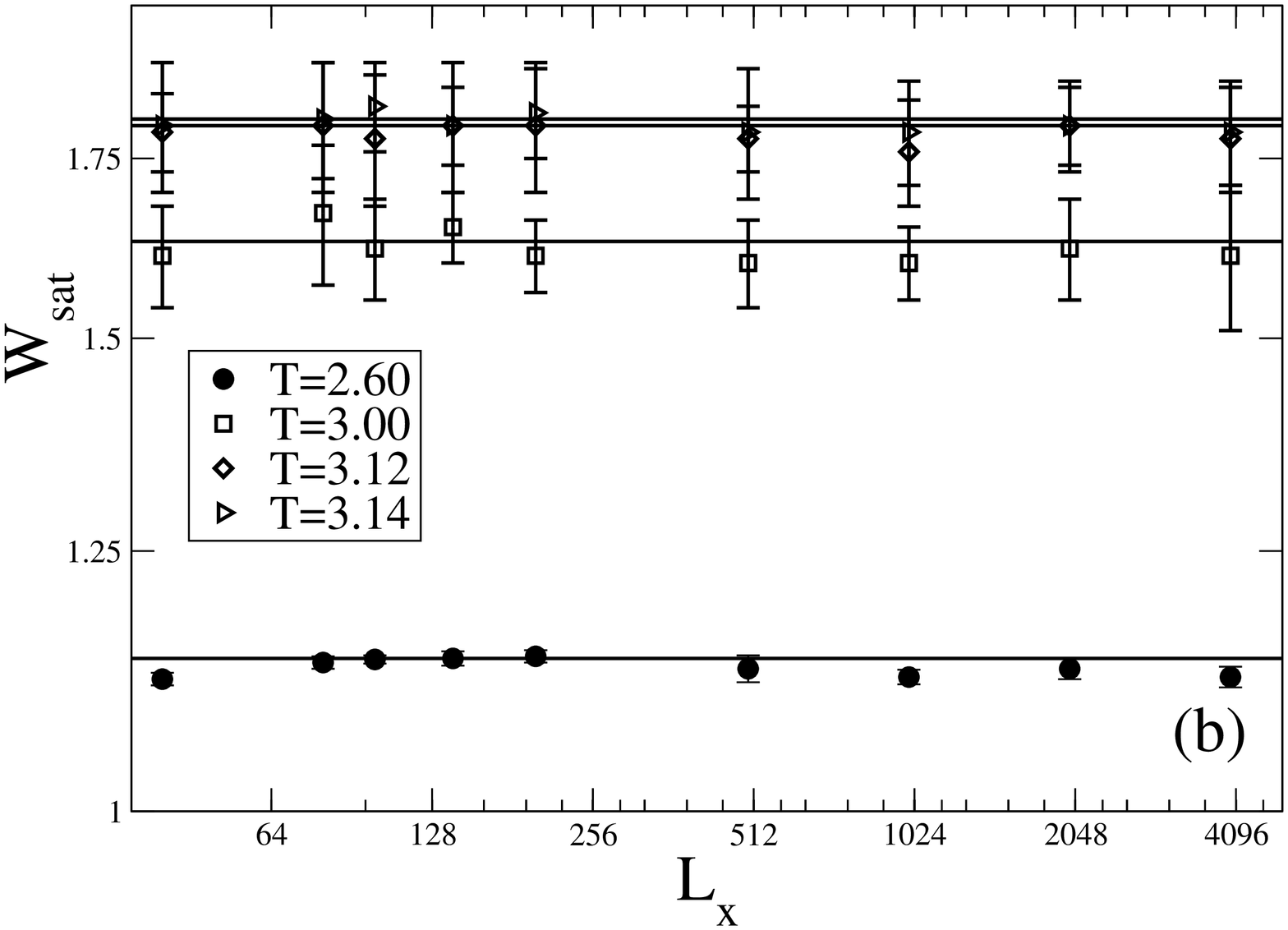}
\caption{Saturation width versus lattice sizes. (a) Linear-log plots of the saturation width $W_{sat}^{2}$ versus
$L_{y}$ obtained for the listed temperatures, including $T=2.05$ $J/k_b$ (see the Introduction). The full lines correspond to regression fits of the measured data by using equation (\ref{fitwsat}) taking $\alpha_I=1/2$; (b) linear-linear plots of  $W_{sat}$ versus $L_x$ and fixed $L_y=20$, for the listed temperatures. The values of $W_{sat}^{2}$ (figure 4 (a)) obtained with $S$ fixed are also included.} \label{W2Lytodo}
\end{figure}

After determining the saturation value of the interface width at different 
temperatures and using lattices of different sizes, we fitted the data of $W_{sat}^{2}$ with the aid of the following Ansatz
\begin{equation}
\label{fitwsat} 
W_{sat}^{2}=A_{1} ln L_{y} + A_{2} L_{y}^{2\alpha_I}, 
\end{equation}
\noindent where the coefficients $A_{i}'s$ are temperature dependent, and the roughness exponent is taken as $\alpha_I = 1/3$ or $\alpha_I = 1/2$, according to the values of $\Delta_B$ determined by Janssen \textsl{et al} or by Garrido \textsl{et al}, respectively (see eq. (\ref{alfadelta}) and the paragraph below it). The plots of the $W_{sat}$ data and the corresponding fits using equation (\ref{fitwsat}) are shown in figure \ref{W2Lytodo} (a) for the case $\alpha_I=1/2$. Figure \ref{W2Lytodo} (a) shows that the Ansatz (eq. (\ref{fitwsat})) fits the saturation values at all temperatures, although the fit is clearly better for lower temperatures, e.g. $T\leq2.96$. This feature is also present if we take $\alpha_I=1/3$ as roughness exponent. Furthermore, figure \ref{W2Lytodo} (b) shows that the saturated width does not depend on either the perpendicular lattice size $L_x$ or the shape factor $S$, since the values of $W_{sat}$ obtained by keeping $S$ fixed are also included in the plot, and practically no difference is noticed from those corresponding to the nonfixed $S$ values. Therefore, we can conclude that our results are not affected by any particular choice of the shape factor. This plot also shows that the interaction between interfaces is negligible since no noticeable changes in $W_{sat}$ are observed when the distance between both interfaces of the single strip is increased due to the change of $L_x$ by keeping the density of particles $\rho_{0}=1/2$ constant.\\

\begin{figure}[H]
\centering
\includegraphics[height=15cm,width=11cm,clip=,angle=270]{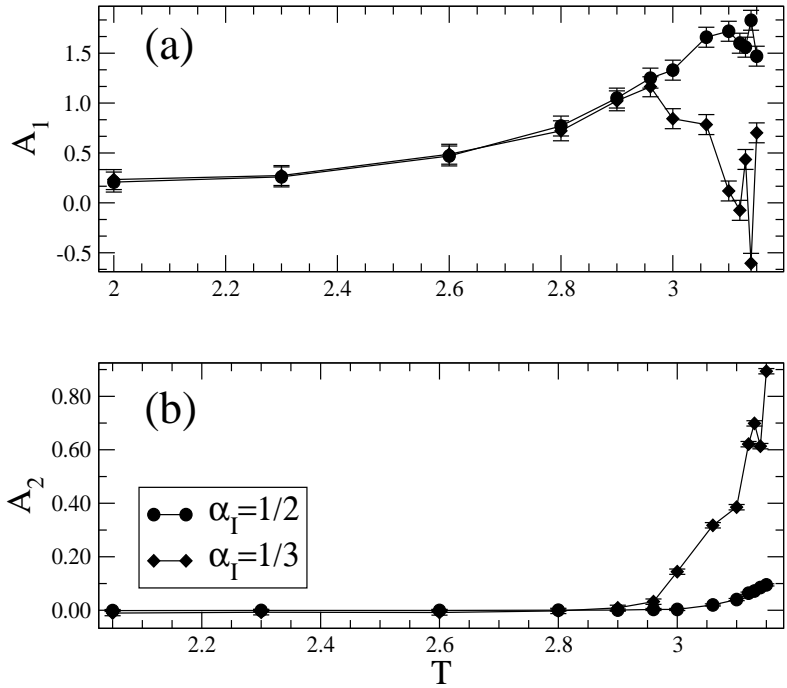}
\caption{(a) and (b) show linear-linear plots of the coefficients $A_1$ and $A_2$ versus the temperature,
 for both cases $\alpha_I=1/3$ and $\alpha_I=1/2$, as indicated in the legend. Data obtained from the fits of 
the data shown in figure \ref{W2Lytodo} according to eq. (\ref{fitwsat}).}
\label{coef}
\end{figure}

The fitting coefficients of equation (\ref{fitwsat}) will help us  
determine which behavior (i.e., logarithmic or power law) becomes relevant along the 
investigated temperature interval. Figure \ref{coef} show plots of $A_1$ and $A_2$ as a function of $T$. 
By comparing figures \ref{coef} (a) and (b), we can observe that in the interval $2.05\leq T \lesssim 2.96$, $A_1$ 
increases monotonically assuming almost the same value for both cases of $\alpha_I$. $A_2$ is practically zero within 
this regime for both cases, indicating that the logarithmic finite-size behavior of $W_{sat}$ dominates over 
the power law. On the other hand, near the bulk critical temperature ($T\ge 3.0$), $A_1$ still grows monotonically and then reaches a saturation value around $A_1\simeq1.75$ for the case $\alpha_I=1/2$, but whereas it fluctuates with a decreasing trend if $\alpha_I=1/3$ is taken. 
However, $A_2$ increases for both tested values of $\alpha$, revealing that the power-law behavior becomes  
relevant for this temperature interval.

\begin{figure}[H]
\centering
\includegraphics[height=15cm,width=11cm,clip=,angle=270]{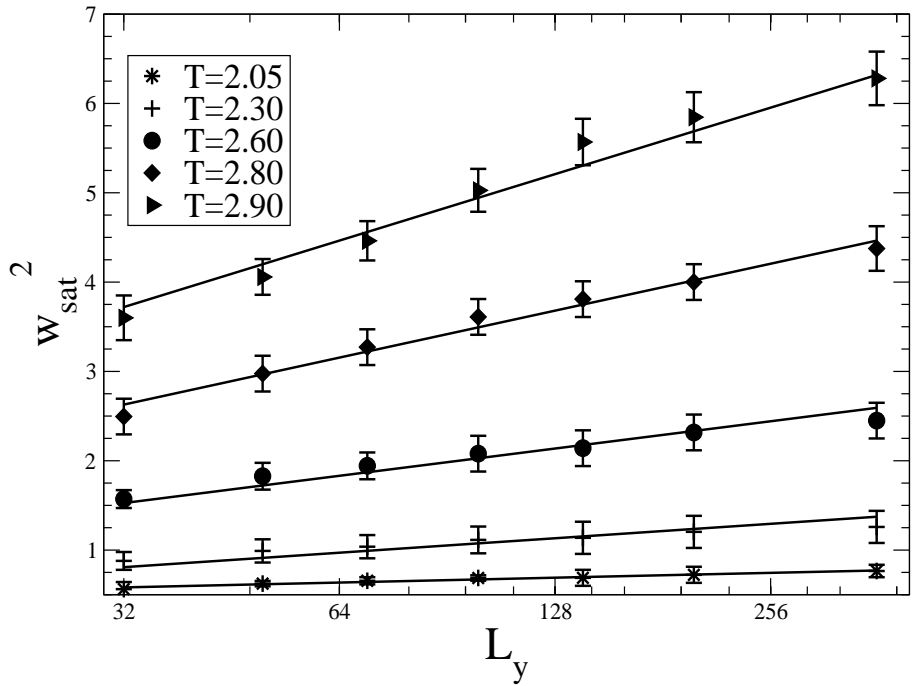}
\caption{Linear-log plot of the saturation values of the roughness
$W_{sat}^{2}$ versus $L_{y}$, for very low temperatures, as
indicated. The data are fitted by the logarithmic term in eq. (\ref{fitwsat}), indicated in the plot by the full
lines.}
\label{wsatl}
\end{figure}

\noindent Since both regimes are clearly distinguishable, we investigated them separatedly. 
In this way, figure \ref{wsatl} shows plots of  $W_{sat}^{2}$ $\propto$ $\ln$ $L_y$ as obtained 
within the low T range. 

\noindent The obtained results show that the behavior $W^2\sim$ $\ln$ $L_y$ holds for all temperatures reported, including $T=2.05$ $J/k_b$, the working temperature used by Leung \textsl{et al} in their simulations \cite{leung2}. This result had already been predicted theoretically by Zia
\textsl{et al.} for the RDLG model \cite{zialog}, and tested by Leung \textsl{et al.} \cite{leung2}. 
However, the best knowledge of ours, 
there are neither theoretical predictions nor numerical tests for the IDLG model 
that we can use for the sake of comparison. These results are consistent with a roughening temperature  for the IDLG model such that $T_R<2.05$.\\ 

\begin{figure}[H]
\centering
\includegraphics[height=15cm,width=11cm,clip=,angle=270]{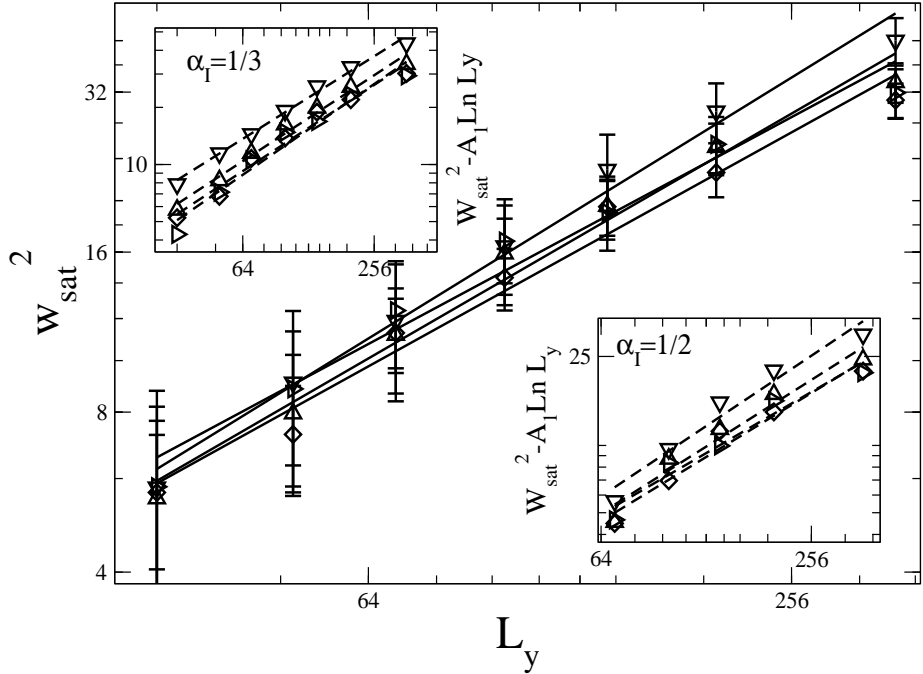}
\caption{Log-log plots of the saturation width $W_{sat}^{2}$
versus $L_{y}$ corresponding to $T=3.12$ ($\lozenge$), $T=3.13$ ($\triangle$), $T=3.14$ ($\triangleright$), and $T=3.15$ ($\bigtriangledown$), respectively. The insets show log-log plots of $W_{sat}^{2}$ corrected by the logarithmic term in equation (\ref{fitwsat}) versus $L_y$ (the error bars are excluded for the sake of clarity). The upper inset shows the corrections and power law fits if $\alpha_I=1/3$ is taken in equation (\ref{wsatkls}), while the lower inset shows the same for the case $\alpha_I=1/2$. In all plots the lines are power-law fits of the data.}
\label{wsatpwr}
\end{figure}

Within the high-temperature interval, the logarithmic growth crosses over  to a power-law 
behavior (see the power-law term of equation (\ref{fitwsat})) according to equation (\ref{wsatkls}). Consequently, an effective roughness exponent $\alpha_{eff}$ can be determined for each temperature. 
Figure \ref{wsatpwr} shows log-log plots of $W^{2}_{sat}$ 
versus $L_y$ obtained for temperatures close to $T_c$. \\

\begin{figure}[H]
\centering
\includegraphics[height=15cm,width=11cm,clip=,angle=270]{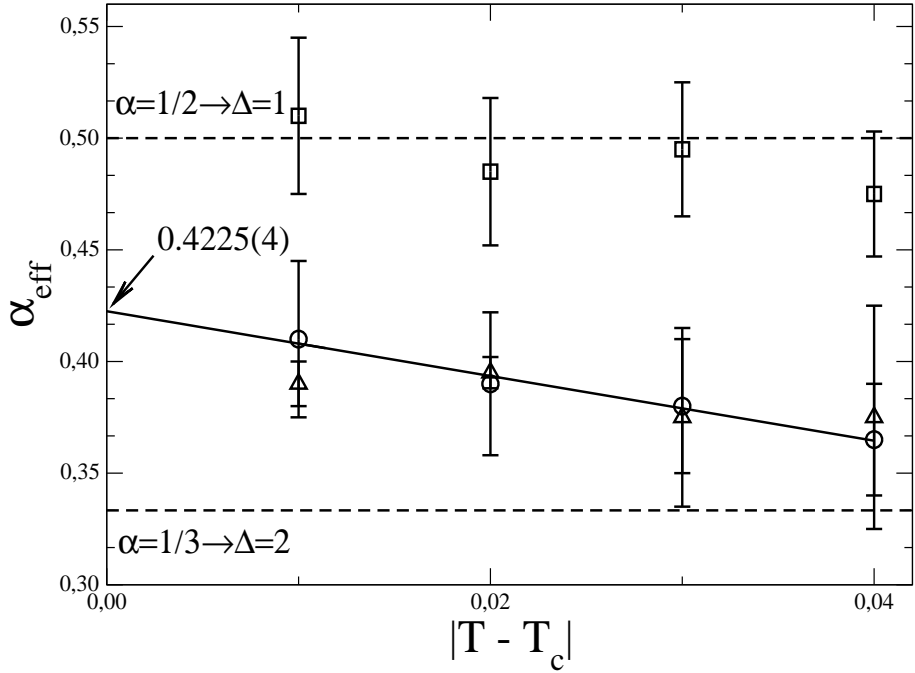}
\caption{Plot of the effective exponent $\alpha_{eff}$ versus 
$|T-T_{c}|$, obtained from the power-law fits of the data shown in the main plot of 
figure \ref{wsatpwr} ($\bigcirc$). The full line is a linear fit of the uncorrected data, 
whose extrapolated ordinate is $\alpha_{eff}(T=T_c)=0.4225(4)$ as it is shown by the arrow.
The dashed lines indicate the predicted theoretical values for $\alpha_I$: the upper line is the value obtained by
taking $\Delta_I=1$, while the lower line shows the value 
of $\alpha_I$ obtained by taking $\Delta_I=2$. Also the values $\alpha_{eff}$ when $W^{2}_{sat}$ is corrected by the logarithmic term in equation (\ref{fitwsat}) are included ($\bigtriangleup$ for $\alpha_I=1/3$, $\square$ for $\alpha_I=1/2$, respectively).}
\label{alfast}
\end{figure}

\noindent In this way, $\alpha_{eff}$ was measured for each $T$, and
plotted as a function of the distance to the critical
temperature. Then, we extrapolated the value of $\alpha_I$
to $T_{c}$. Our results are displayed in figure \ref{alfast}. 
According to the trend observed in the plot, it is expected
that $\alpha_{eff} \rightarrow 1/2$ when $T\rightarrow T_c$, since all the determined values are  
over $\alpha_I= 1/3$ (lower dashed line in figure \ref{alfast}), calculated by using $\Delta_I=2$ in equation 
(\ref{alfadelta}). A linear extrapolation of the data gives 
$\alpha_{eff}(T=T_c)=0.4225(4)$, which is close to the value expected for $\alpha_I$ if it is calculated by using 
$\Delta_I=1$. However, the data $\alpha_{eff}$ can be improved if the logarithmic term in equation 
(\ref{fitwsat}) is substracted from $W_{sat}^{2}$. By fitting the corrected values of $W_{sat}^2$ with equation (\ref{wsatkls}) (see insets of figure \ref{wsatpwr}), 
we observe that all values are, again, well above $\alpha_I=1/3$ (indicated in the plot by the lower dashed line) 
calculated by using $\Delta_I=2$ (see equation (\ref{alfadelta})). The corrected values of $\alpha_{eff}$ considering the coefficients $A_1$ and $A_2$ obtained by fitting the data with the aid of equation (\ref{fitwsat}) by taking  
$\alpha_I=1/3$, are not significantly improved with respect to the uncorrected values. In contrast, by using the coefficients obtained for $\alpha_I=1/2$, the corrected values of $\alpha_{eff}$ are markedly improved and lie close to $\alpha_I=1/2$ (indicated in the plot by the upper dashed line). In view of these 
results, we can conclude that $\alpha_{eff}$ approaches $\alpha_I=1/2$ in the $T\rightarrow T_{c}$ limit. 
This is essential for the clarification of the issue of the universality class of the IDLG model. By using
equation (\ref{alfadelta}) it is possible to compute the anisotropic 
exponent, which gives $\Delta_I \approx 1$. Since we assumed in equation (\ref{alfadelta}) that the anisotropic relation found for the bulk critical behavior holds also for the interfacial behavior, we can conclude that this result is in agreement with the value predicted by Garrido \textit{et al} on analyzing an
alternative mesoscopic equation \cite{gallegos1}. 

\begin{figure}[H]
\centering
\includegraphics[height=15cm,width=11cm,clip=,angle=270]{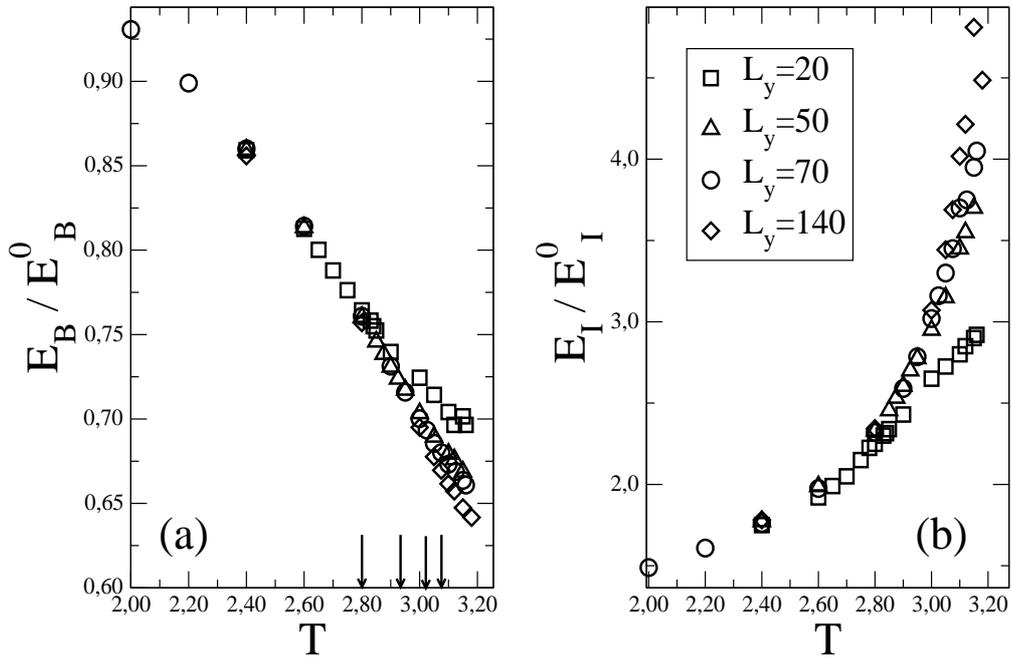}
\caption{(a) and (b) show plots of the bulk and interface normalized NESS energies corresponding to different sizes $L_y$ (indicated in the legend) versus the temperature $T$. Here $E^{0}_{B}$ and $E^{0}_{I}$ are the energies of the initial GSC configuration. The change of convexity in the plots of figure 9 (a) is indicated by the arrows on the $T$ axis. }
\label{enerbi}
\end{figure}

\begin{figure}[H]
\centering
\includegraphics[height=15cm,width=11cm,clip=,angle=270]{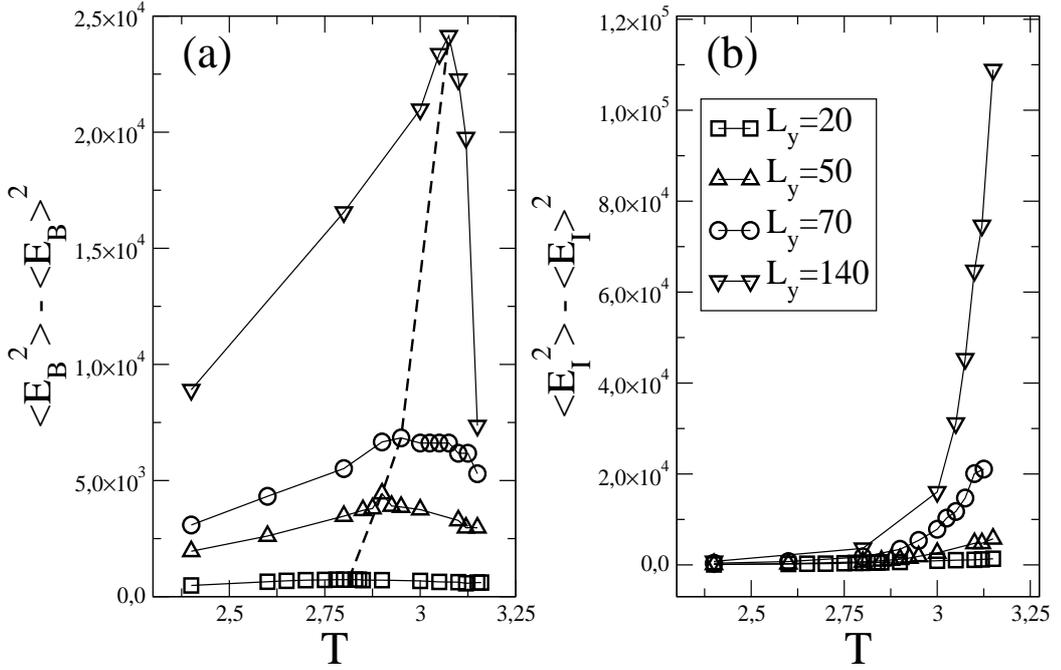}
\caption{Plots of the bulk energy (a) and interface energy (b) fluctuations. The dashed line in figure (a) indicates the shift of the maxima as a function of $L_y$. }
\label{fluct}
\end{figure}

The behavior of $W_{sat}^{2}$ for $T\gtrsim$ $3$ can be further investigated by studying the bulk and interface ``energies'', given by equation (\ref{Hamil}), and their fluctuations, defined as 
\begin{equation}
\label{eifluct} 
\sigma_{B,I}=\langle E^{2}_{B,I}\rangle - \langle E_{B,I}\rangle ^{2},
\end{equation}
\noindent where $E_{B,I}$ is the bulk or interface energy (subscripts $B$ or $I$, respectively). The obtained results are displayed in figure \ref{enerbi} (a) and (b), respectively. 
As the temperature increases, the bulk energy (figure \ref{enerbi} (a)) decreases due to the presence of holes in the bulk and exhibits a change in the convexity for $2.8\lesssim T \lesssim 3.1$, as indicated by the arrows in the figure. This behavior is similar to that observed in the equilibrium 2d Ising model. In contrast, the interface energy grows monotonically due to the interface roughening. \\
On the other hand, figure \ref{fluct} (a) shows that for all system sizes parallel to the interface ($L_y$) the fluctuations of the energy as a function of $T$, exhibit a peak. However, the interface energy fluctuations still grow when both the system size and temperature increase. These phenomena, together with the (non-) convexity change in the (interface) bulk energy, indicate us that the change in the behavior of $W_{sat}^{2}$ with $T$ is in fact a crossover to the bulk critical behavior. These findings also justify the assumption that the anisotropy of the interface and bulk would be the same (eqs. (\ref{zbdkls}), (\ref{zbeate}) and (\ref{alfadelta})).

\begin{figure}[H]
\centering
\includegraphics[height=15cm,width=11cm,clip=,angle=270]{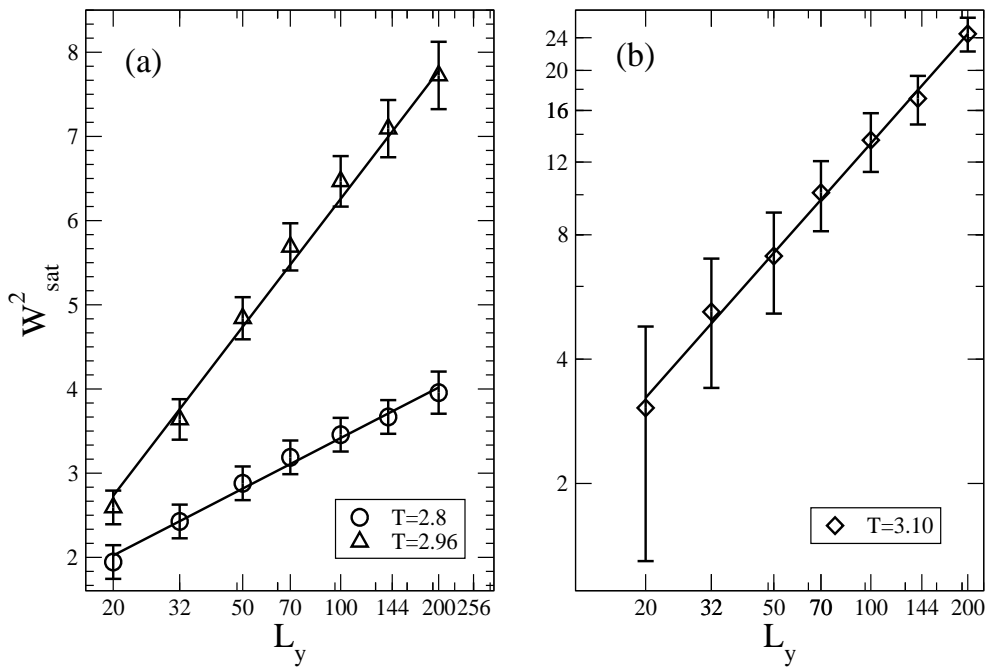}
\caption{Data corresponding to the IRDLG model obtained at
different temperatures, as indicated.
(a) Linear-log plots of $W^2_{sat}$ versus $L_y$.
The full lines are the fit of numerical data
obtained by considering the logarithmic term of equation (\ref{fitwsat}) only.
(b) Log-log plot of $W^2_{sat}$ versus $L_y$.
In this case, the numerical data are fitted
to a power law $W^{2}\propto L_{y}^{2\alpha_{I}}$,
according to eq. (\ref{wsatkls}), or to the power-law term in eq. (\ref{fitwsat}).}
\label{detall}
\end{figure}

We also performed test simulations of the IRDLG model.
In fact, figures \ref{detall} (a) and (b) show plots of the
saturation width versus $L_{y}$ obtained for three different 
temperatures and using the same lattice sizes as in the case of the 
IDLG model. These figures qualitatively show the same behavior 
already observed for the IDLG model (see figure \ref{W2Lytodo}), that 
is,
the saturation width grows logarithmically with $L_y$ at low
temperatures, in agreement with the theoretical results
of Zia \textsl{et al.} \cite{zialog} and the numerical simulations of Leung \textsl{et al.} \cite{leung2}.
Subsequently, the interface width  crosses over to a power-law 
behavior for $T\lesssim T_c$.

From the fit of the data shown in figure \ref{detall} (b), the estimated value of the roughness is $\alpha_{eff}=0.44(2)$. 
This result is not only in agreement with the fact that
$\alpha_I=1/2$ since one has $\Delta_I=1$ in equation (\ref{alfadelta}) for the IDLG model, but also validates our scaling approach in Section \ref{3} for the interface roughness for the IRDLG model. However, we recognize that, in figure \ref{detall} (b), the data have been fitted within one decade only, because either the strip used as initial condition no longer percolates, or eventually it splits into more than one strip.\\

At this point, and before continuing with the description of our results concerning the dynamic critical behavior of the interface, we would like to stress that the above-summarized results exhibit some discrepancies with those previously obtained by using Monte Carlo simulations by Leung \textit{et al} \cite{leung2}. Our results showed that for both the IDLG and IRDLG models the roughness grows logaritmically with the longitudinal size $L_y$  at low temperatures $T\ll T_c$, $W^2 \sim \ln L_y$, while near the critical point $T\lesssim T_c$ it increases as a power-law $W^2\simeq L_{y}^{\alpha_I}$. On the other hand, Leung \textit{et al} found that the roughness is supressed for both the DLG and RDLG models at low temperatures ($T\ll T_c$) and large lattices \footnote{A brief but detailed summary of these results is included in the Introduction}.
Although we cannot give a definitive answer to explain the origin of these differences, we can state a few possible assumptions, which may deserve further research in the future: 1) The coarse grain method. In order to compute the interface position $h(j,t)$ (see eq. (\ref{hmediodef})), Leung \textit{et al} considered a method of smoothing the interface by applying a coarse graining process, which consists of the distribution of the occupation number of each site (0 or 1) between its neighbors, assigning 0 or $1/5$ to each neighbor and to the site itself, respectively. When the algorithm is iterated, the resulting $h(j,t)$ is a single-valued function of $j$ that takes into account only the fluctuations due to the capillary waves and not the contribution of bubbles and overhangs. Perhaps, by using Leung's \textsl{et al} procedure, the saturated interface width becomes oversmoothed and simplified to observe a logarithmic increase. In fact, we still find a logarithmic increase of $W_{sat}^{2}$ as a function of $L_y$ in our simulations at the same temperature and field magnitude they reported, namely $T=2.05$ $J/k_b$ and $E=50$, respectively. 2) The error in the computation of the exponent $\eta\sim1.33$ of the structure factor $G(q)\sim q^{-2+\eta}$ (small $\overrightarrow{r}$) for the DLG model was not reported. This issue is important, because if $\eta$ has a large error, one can no longer give any conclusion about the nature of the asymptotic behavior in the $q\rightarrow0$ limit, and consequently to distinguish a logarithmic or a power-law behavior of the interface width \cite{zialog}. 3) The temperature range. Regrettably, the results of Leung \textsl{et al} for both the DLG and the RDLG models were obtained for a single temperature ($T=2.05$ $J/k_b$), in contrast to the wide range of T covered by our simulations. By using different temperatures we are able to clearly observe the logarithmic dependence of the interface width, namely $W_{sat}^{2}=A_1 \ln L_y$ (figure \ref{coef}). Since $A_1$ is vanishing when $T$ is decreased ($2.05\leq T \leq2.96$), it would be very difficult to distinguish a logarithmic dependence just by measuring the interface at a single and rather low $T$ ; 4) The finite size. In their work, Leung \textit{et al} used lattices of maximum longitudinal size $L_y=600$ lattice units. However, their results were not conclusive to explain the behavior in the $L_y\rightarrow\infty$ limit since the data for the RDLG seem to exhibit a gap in the $q\rightarrow0$ limit. On the other hand, our largest lattice size used was $L_y=360$ lattice units, which demands a lot of computational time due to the algorithm that identifies the interface, the long time required to reach the saturation of the width, and the large number of realizations that one needs to obtain good statistics.\\
It is also worth mentioning that our algorithm has succesfully been tested twice under different circumstances: a) for the characterization of the interface of diffusion fronts, which have many bubbles and overhangs. In this case the numerical results are fully consistent with the predictions of the well accepted standard percolation theory \cite{vero}. b) For the characterization of the interface between magnetic domains of the Ising model (see figure \ref{e0} of the present paper and related discussion). \\

\begin{figure}[H]
\centering
\includegraphics[height=15cm,width=11cm,clip=,angle=-90]{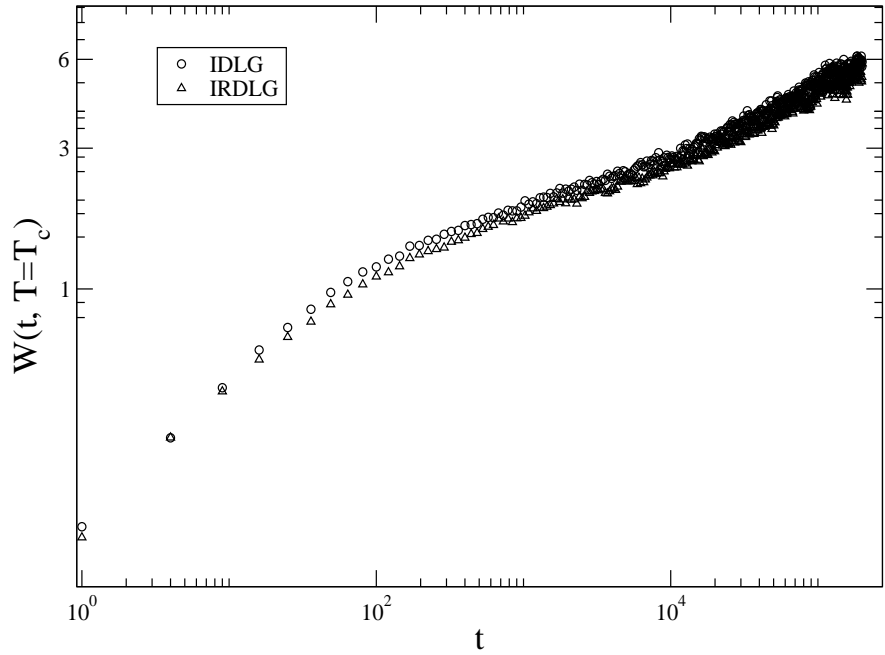}
\caption{Log-log plots of the interface width $W$ versus $t$
for both the IDLG and IRDLG models, measured at their 
respective critical temperatures.
Results obtained using lattices of size $L_x = 346$ and $L_y=524$.}
\label{2modelos}
\end{figure}

In order to investigate the critical interfacial properties of both 
the IDLG and the IRDLG models, we also measured the time
dependence of the interface width at $T_{c}$, as shown 
in figure \ref{2modelos}. It is worth mentioning that both models exhibit the same behavior at criticality, as evidenced by the overlap of the numerical data shown in figure \ref{2modelos}.

\begin{figure}[H]
\centering
\includegraphics[height=15cm,width=11cm,clip=,angle=270]{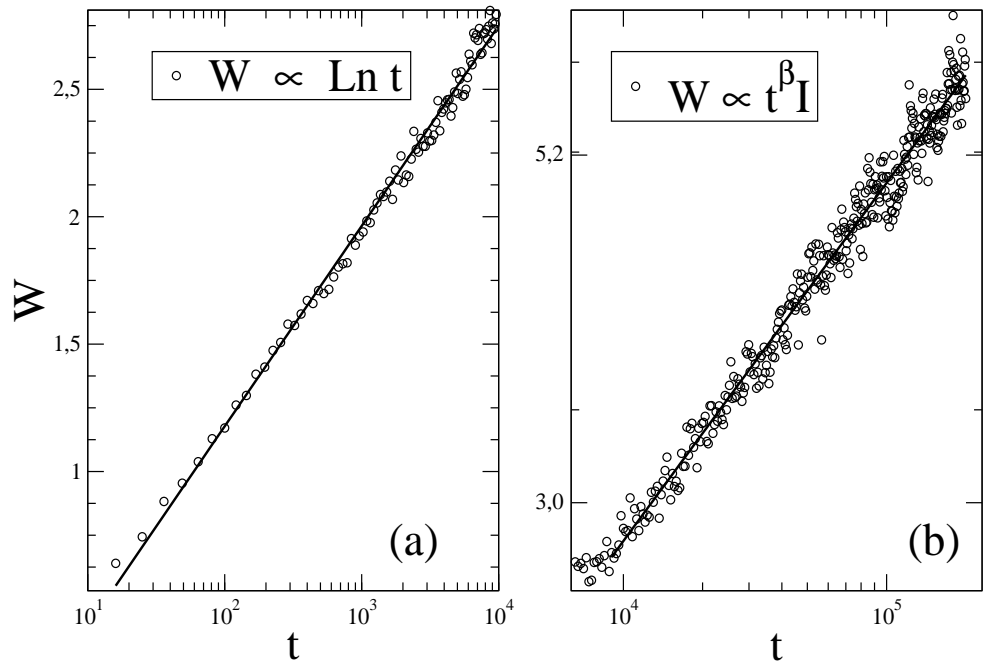}
\caption{Growth regimes observed during the time evolution of
the interface width in the IDLG model at criticality.
(a) Logarithmic growth; (b) Power-law growth.}
\label{regikls}
\end{figure}

\begin{figure}[H]
\centering
\includegraphics[height=15cm,width=11cm,clip=,angle=270]{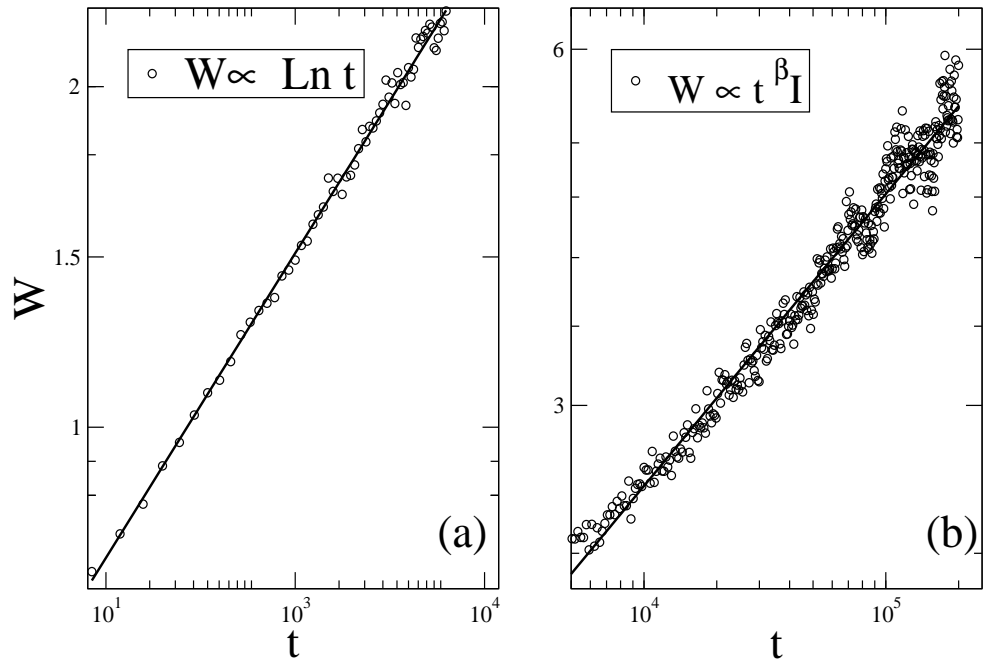}
\caption{Growth regimes observed during the time evolution of
the interface width in the IRDLG model at criticality.
(a) Logarithmic growth; (b) Power-law growth.}
\label{regirkls}
\end{figure}

Getting further inside the time evolution of the width, two well
defined growth regimes can be distinguished. At early times the 
growth is logarithmic $W \propto$ $\ln (t)$, and it lasts almost three
decades ($10-10^{4}$ MCS) for both models, see figures \ref{regikls}(a) 
and \ref{regirkls}(a). On the other hand, at later times, a
crossover to a power-law growth regime is observed, $W\propto t^{\beta_{I}}$, as expected from equation (\ref{diverperp}). 
\newpage

\newpage

Focusing on the power-law behavior, we can make a fit of the data
to compute the growth exponent of the interface, which yields
$\beta_{I}(IDLG)=0.245(2)$, so that  $z_{\perp}^{I}(IDLG) =4.08(4)$
according to equation (\ref{diverperp}). 
On the other hand, by fitting the power-law divergence of the interface
width for the IRDLG model, the growth exponent $\beta_{I}(IRDLG) =
0.247(2)$ is obtained. Therefore, $z_{\perp}^{I}(IRDLG) = 4.05(3)$. These values of the exponents strongly suggest that $z_{\perp}^I=z_{\perp}^B$ for both models. Also, as expected, the determination of the dynamic exponent $z_{\perp}^I$ is not helpful in assigning the universality
class of the studied models since, within error bars, theoretical calculations and numerical results are consistent
with $z_{\perp}^B \simeq 4$, for both models and approaches. In any case, these results further validate our scaling approach for the interface roughness.

\section{Conclusions}
\label{5}
In this work we studied the dynamic and steady-state behavior of the strip-gas interfaces in both IDLG and IRDLG models. We measured the average width and the roughness exponent $\alpha_{I}$ for a wide range of temperatures $T\leq T_c$, as a function of the lattice sizes and the shape factor.
Starting with the IDLG model, the interface width $W^{2}$ reaches a saturation value $W^{2}_{sat}$ that depends on the lattice size parallel to the external field direction, $L_y$ (figure \ref{W2Lytodo} (a)), but it is independent of both the transversal size and the shape (figure \ref{W2Lytodo} (b)). In the low-temperature regime, $T\lesssim3.0$, $W^{2}_{sat}$ exhibits a logarithmic dependence on $L_y$, as has been theoretically found by Zia \textsl{et al.} \cite{zialog} and tested numerically in a subsequent work by Leung \textsl{et al.} \cite{leung2} for the RDLG model. This behavior is consistent with the presence of a roughening transition at very low $T$, presumably at $T=0$ as in the RDLG model \cite{zialog}.
Then, close to the critical zone, $3.00\lesssim T \leqslant T_c=3.20$, $W^{2}_{sat}$ crosses over to a power law of the form $W^{2}_{sat}\propto L_{y}^{2\alpha_{eff}}$. The value of $\alpha_{eff}$ is estimated to be $\alpha_{I}=\alpha_{eff}(T\rightarrow Tc)\approx 1/2$ so $\Delta_I \simeq 1$ using equation (\ref{alfadelta}). Since we assume that the bulk anisotropic behavior holds also for the interfacial behavior, this value is in accordance with that computed by Garrido \textsl{et al.} for the \textsl{bulk} anisotropic exponent $\Delta_B$. Furthermore, the field-theoretical predictions of Janssen \textsl{et al} for this value ($\Delta_B=2$) seems to be ruled out.\\

In order to investigate the change of behavior in $W^{2}_{sat}$ around $T=3.00$, we measured the bulk and interface energies, together with their respective fluctuations. The curves of the bulk energy versus the temperature show a change in convexity around the size-dependent critical temperature. The energy fluctuations are peaked at the finite-size critical temperature, and they shift to the $L_y\rightarrow\infty$ critical temperature when $L_y$ is increased. In contrast, the interface energy and its fluctuations grow monotonically as the temperature and the lattice size are increased. These facts allow us to conclude that the change in the behavior of $W^{2}_{sat}$ around $T=3.00$ is in fact a crossover to the bulk critical behavior, and apparently it is not related to any interface phenomena such as a roughening transition.\\
Later, the time evolution of the interface width at the critical point was measured. In this case $W$ does not saturate and exhibits two growing stages. In the first stage, $W$ grows logarithmically for several decades, $W\propto$ $ln$ $t$, and in the following stage $W$ grows obeying a power-law, $W\propto t^{\beta_{I}}$, with growth exponent $\beta_{I}=1/z_{\perp}^I$. The obtained value of the dynamic exponent was $z_{\perp}^I=4.08(4)$, which is in good agreement with the theoretically estimated values for the bulk critical behavior, that is  $z_{\perp}^B\approx 3.996$ \cite{gallegos1} and $z_{\perp}^B\approx 4$ \cite{JS}.\\ 
On the other hand, the same studies performed for the IRDLG model gave practically the same results. At low temperatures $W^{2}_{sat}\propto ln L_y$, confirming the predictions of Zia \textsl{et al.} \cite{zialog} about the existence of a roughening transition at $T=0$. Then, $W^{2}_{sat}$ crosses over to a power law for $T\lesssim T_c$, with an effective roughness exponent that is similar to the one measured in the IDLG model at the same temperature. At the critical point, the time evolution of $W$ is the same as in the IDLG model, exhibiting the same logarithmic and power-law growth regimes. In this last stage the dynamic exponent of the evolution was obtained giving  $z_{\perp}^I=4.05(3)$, in agreement with the estimated value of $z_{\perp}^B$ for the IDLG model \cite{beate, gallegos1}.\\

In short, we think that our study is valuable because it exploits the saturating character of the interface in the $T < T_c$ interval, extracting useful information about the interface behavior relating it to the bulk critical behavior. Furthermore, our generalization of the dynamic scaling approach of Family and Vicsek to the driven lattice gas allows for an independent estimation of the anisotropic exponent, further estimulating the discussion about the universality class of the model.
 
\section{Acknowledgments}This work was made under the support of CONICET, UNLP, 
and ANPCyT (Argentina). Discussions with M. A. Mu\~{n}oz are also acknowledged.


\begin{thebibliography}{1}

\bibitem{beate} B. Schmittmann and R. K. P. Zia, 
{\it Statistical Mechanics of Driven Diffusive System}, 
in Phase transitions and Critical Phenomena, Vol. 17, 
edited by C. Domb and J. Lebowitz (Academic London, 1995).
\bibitem{priv}V. Privman (ed.), in: {\it Nonequilibrium Statistical Mechanics in One Dimension}
(Cambrige University Press, 1997)
\bibitem{marr}J. Marro and R. Dickman, in: \emph{Nonequilibrium Statistical Mechanics of Lattice
Models} (Cambridge University Press, 1999)
\bibitem{kls} S. Katz, J. L. Lebowitz and H. Spohn, Phys Rev. B {\bf 28}, 1655 (1983).
\bibitem{pablo} P. I. Hurtado, J. Marro, P. L. Garrido and E. V. Albano, Phys Rev. B {\bf 67}, 014206 (2003).
\bibitem{JS} H. K. Janssen and B. Schmittmann, {\em Z. Phys.\/} B {\bf 
64}. 503
(1986). K. -t. Leung and J. L. Cardy, J. Stat. Phys. {\bf 44}, 567
(1986); ibid {\bf 45}, 1087 (1986) (Erratum).
\bibitem{ktl} K. -t. Leung, Phys. Rev. Lett. {\bf 66}, 453 (1991).
\bibitem{rkls} B. Schmittmann and R. K. P. Zia, Phys. Rev. Lett. {\bf 66}, 357 (1991).
\bibitem{gallegos1} P. L. Garrido, F. de los Santos and M. A. Mu\~noz,
Phys. Rev. E {\bf 57}, 752 (1998).  P. L. Garrido, M. A. Mu\~noz and F. de los Santos,  Phys. Rev. E. {\bf 61}, R4683 (2000); F. de los Santos, P. L. Garrido and M. A. Mu\~noz, Physica A {\bf 296}, 362 (2001).
\bibitem{tanos} S. Caracciolo, A. Gambassi, M. Gubinelli and A.
Pelissetto, Phys. Rev. Lett. \textbf{92}, 029601 (2004).
\bibitem{reply} E. V. Albano and G. Saracco, Phys. Rev. Lett \textbf{92}, 029602 (2004).
\bibitem{gallegos2} A. Achahbar, P. L. Garrido, J. Marro and M. A. Mu\~noz,
Phys. Rev. Lett. {\bf 87}, 195702 (2001).
\bibitem{alsa} E. V. Albano and G. Saracco, Phys. Rev. Lett. {\bf 88}, 145701 (2002).
\bibitem {gamba}S. Caracciolo, A. Gambassi, M. Gubinelli and A.
Pelissetto, J. Phys. A: Math. Gen. \textbf{36}, L315 (2003); S. Caracciolo, A. Gambassi, M. Gubinelli and A.
Pelissetto, J. Stat. Phys. \textbf{115}, 281 (2004). S. C. Park and J.-M. Park, J. Kor. Phys. Soc. \textbf{49} No. 4, 1365, (2006).
\bibitem{gamba2} S. Caracciolo, A. Gambassi, M. Gubinelli and A.
Pelissetto, Phys. Rev. E \textbf{72}, 056111 (2005).
\bibitem{bara} A. L. Barabási and H. E. Stanley, 
in \textit{Fractal Concepts in Surface Growth}, 
Cambridge University Press, New York, 1995.
 
\bibitem{lapu} J. Lapoujade, Surf, Sci Rep. \textbf{20}, 191 (1994).
\bibitem{weeks} J. D. Weeks and G. H. Gilmer, Adv. Chem. Phys \textbf{40}, 157 (1979).
\bibitem{bruk} E. Br\"ukner and D. Stauffer, Z. Phys. B \textbf{53}, 241, (1983); K. K. Mon, S. Wansleben, D. P. Landau and K. Binder, Phys. Rev. B \textbf{39}, 7089 (1989).
\bibitem{vanB} see for example H. van Beijeren and I. Nolden, {\it Structure and dynamics of Surfaces II} (eds. W. Schommers and P. von Blackenhagen). {\it Topics in Current Physics}, Vol 43, Springer, Berlin, 1987.
\bibitem{bab} D. B. Abraham and P. Reed, Phys. Rev. Lett. {\bf 33}, 377 (1974).
\bibitem{leung} K. -t. Leung, J. Stat. Phys. \textbf{50}, 405 (1988).



\bibitem{mon} K. -t.  Leung, K. K. Mon, J. L Vall\'es and R. K. P. Zia, Phys. Rev. Lett. {\bf 61}, 1744 (1988); Phys. Rev. B \textbf{39}, 9312 (1989).

\bibitem{leung2}K. -t. Leung and R. K. P. Zia, 
J. Phys A: Math.  Gen. \textbf{26}, L737 (1993).
\bibitem{zialog} R. K. P. Zia and K -t Leung, 
J. Phys. A: Math. Gen. \textbf{24}, L1399 (1991).
\bibitem{famvis} F. Family and T. Vicsek, Phys. A \textbf{18}, L75 (1985).
\bibitem{turski} L. A. Turski and J. S. Langer, Phys. Rev. A {\bf 22}, 2189 (1977).
\bibitem{kawa} K. Kawasaki and T. Ohta, Prog. Theo. Phys. {\bf 68}, 129 (1982).
\bibitem{yeung}C. Yeung J. L. Mozos, A. Hernández- Machado and D. Jasnow, J. Stat. Phys. \textbf{70}, 1149 (1993).
\bibitem{hav} J. -F. Gouyet, M. Rosso and B. Sapoval, in \textit{Fractals and Disordered Systems}, edited by A. Bunde and  S. Havlin, Springer-Verlag, Berlin, 1991; J. Kertész and T. Vicsek, in \textit{Fractals in Science}, edited by A. Bunde and  S. Havlin, Springer-Verlag, Berlin, 1995.

\bibitem{vero} E. V. Albano and V. C. Chappa, Physica A \textbf{327}, 18 (2003); V. C. Chappa and E. V. Albano, J. Chem. Phys. {\bf 121}, 328 (2004).




\end{thebibliography}
\end{document}